\documentclass[reprint,amssymb, amsmath, aip]{revtex4-1}
\usepackage{graphicx}
\usepackage{float}
\usepackage{color}
\usepackage{epsfig}
\usepackage{cleveref}
\usepackage{docs}%
\usepackage{bm}%
\usepackage[colorlinks=true,linkcolor=blue]{hyperref}%
\expandafter\ifx\csname package@font\endcsname\relax\else
 \expandafter\expandafter
 \expandafter\usepackage
 \expandafter\expandafter
 \expandafter{\csname package@font\endcsname}%
\fi
\begin{document}
\title{Effect of magnetic field on the hysteresis phenomena and floating potential oscillations in a reflex plasma source}
\author{R. Rane}
\email{ramu@ipr.res.in}
\address{Institute For Plasma Research, Bhat, Gandhinagar,Gujarat, India, 382428}%
\affiliation{ Homi Bhabha National Institute, Training School Complex, Anushakti Nagar, Mumbai 400085, India}%
\author{P. Bandyopadhyay}%
\address{Institute For Plasma Research, Bhat, Gandhinagar,Gujarat, India, 382428}%
\author{M. Bandyopadhyay}%
\address{Institute For Plasma Research, Bhat, Gandhinagar,Gujarat, India, 382428}%
\author{Gopi Kishan Sabavath}%
\address{Birla Institute of Technology, Mesra, Ranchi, India-835215}%
\author{A. N. Sekar Iyengar}%
\address{Saha Institute of Nuclear Physics, Kolkata, India-700064}%
\author{S. Mukherjee}%
\address{Institute For Plasma Research, Bhat, Gandhinagar,Gujarat, India, 382428}%
%\date{\today}
%#####################################################################################
\begin{abstract}
An experimental investigation on the periodic and chaotic oscillations in a reflex plasma source in presence of magnetic field is presented. The experiment is conducted in a reflex plasma source, consisting of two cathodes and a ring anode. A penning type DC glow discharge in an uniform axial magnetic field is initiated in the background of argon gas. The current-voltage characteristics near the breakdown voltage show a hysteresis with two distinct discharge current regimes. The effect of magnetic field on the discharge current and floating potential oscillations is studied when the discharge is operated within this hysteresis loop. At a typical axial magnetic field, the discharge transits from high discharge current regime (beyond 4--5~mA), an oscillation free regime, to a low discharge current regime (less than 1~mA). Depending upon the discharge parameters, low discharge current regime shows either the periodic or chaotic oscillation in the frequency range of 1-50~KHz. The frequency of periodic oscillation increases with the increase in magnetic field up to 90~Gauss and with further increase in magnetic field, the periodic oscillation becomes chaotic in nature. 
\end{abstract}
%\submitto{\NJP}
\maketitle
\section{Introduction} 
The hysteresis in a physical, biological and engineering systems is a parametric dependency of the state on its history, which shows a clear signature of nonlinearity in a system. This nonlinear behaviour can be prominent near the transition region. An extensive study on hysteresis phenomena in discharge current-voltage characteristics has been carried out during the last few decades \cite{merlino1984,bosch1986,timm1992,bosch1986b}. The transition in the discharge current mainly occurs due to the change of different discharge parameters such as discharge voltage, neutral gas pressure and electrode configuration for both magnetised \cite{merlino1984,bosch1986} and unmagnetised\cite{timm1992,bosch1986b} plasmas. In case of magnetised plasmas, the magnetic field can also acts as a threshold parameter for this transition. Under certain conditions with filamentary or hot cathode, the discharge exhibits the  above hysteresis phenomena along with the low frequency oscillations in the low current region\cite{merlino1984,bosch1986,timm1992,bosch1986b}. This self-oscillations in the low current regime of a hysteresis loop in thermionic discharges is explained using nonlinear dynamics by Greiner et al.\cite{greiner1993,greiner1995,klinger1995}, whereas the effect of the different discharge parameters on these self-oscillation is later studied by Ding et al.\cite{ding1996}. In an another simulation\cite{lee1998} work, three different modes i.e. anode glow mode (AGM), temperature limited mode (TLM) and double layer mode (DLM) have been discussed in beam driven collisional discharge plasma. Recently, Cho has given an alternative explanation to the mechanism behind self-sustained oscillations of ions in dc glow discharges and dusty plasma\cite{cho2013}. Stable and unstable discharge mode in thermionic gas discharge has been studied experimentally and the results are then compared with the results of  particle in cell (PIC) simulation\cite{klostermann1994} studies. Very recently, a comparative study of nonlinear dynamics of magnetised and un-magnetised glow discharge plasma has also been studied by Sharma et al.\cite{sharma2013}. The hysteresis, associated oscillations and their nonlinear dynamics e.g. chaotic plasma behaviour and period doubling are studied in a cylindrical electrode system\cite{nurujiaman2007,kumar2014}.\par
%#######################################################################
\begin{figure*}[!]
\includegraphics[width=0.90\textwidth]{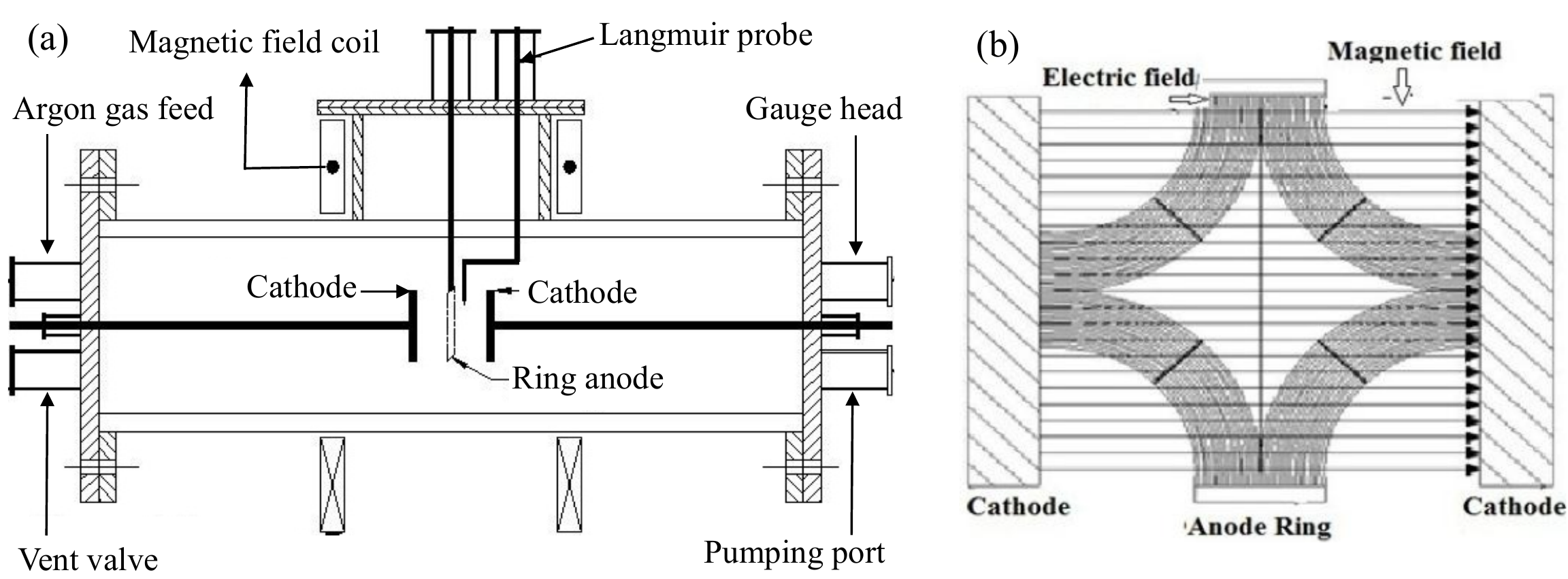}
\caption{Schematic of (a) experimental setup and (b) typical vacuum electric and magnetic field profiles.}
\label{figure1}
\end{figure*}
%########################################################################
In this experimental investigation, we report the hysteresis phenomenon and the associated self-oscillations in a reflex glow discharge plasma. A reflex plasma source with two planer cathodes and a ring anode (placed in between the cathodes) with external applied magnetic field is termed as one of the modified glow discharge This source works as a hybrid between hollow cathode discharge and penning discharge. Due to the external magnetic field, the hollow cathode effect gets enhanced at particular inter-cathode distance, which prevents the electron loss in the radial direction. In this kind of discharge, electrons are emitted perpendicular to cathode and move like an electron beam parallel to magnetic field. The reflex plasma source with high plasma density and low electron temperature have been studied  thoroughly by Toader \textit{et. al.}\cite{toader2004} . Later on, this type of plasma source has been characterised for different operating parameters, e.g. filled pressure, gas flow and magnetic field\cite{toader2005,toader2000}. \par
The effect of the magnetic field on the hysteresis characteristics near the breakdown voltage associated with low frequency oscillations has been sparsely explored in reflex plasma source. This paper reports the detailed study on the hysteresis phenomenon and associated nonlinear oscillation in the reflex plasma source having magnetically constricted anode ring. \par
The rest of the paper is organised as follows: in the next section (Sec.~\ref{sec:exp}), a detailed description of the experimental set-up will be the provided. The result on the hysteresis phenomenon and associated oscillations will be discussed in Sec.~\ref{sec:results}. A concluding remark will be given in Sec.~\ref{sec:conclusion}.
%%%%%%%%%%%%%%%%%%%%%%%%%%%%%%%%%%%%%%%%%%%%5
\section{Experimental  set-up and diagnostic}\label{sec:exp}
The experiment is performed in a hollow cylindrical glass discharge tube of 200 mm diameter and 300 mm length. The schematic diagram of experimental set up is shown in Fig.~\ref{figure1}(a). The glass chamber has several axial and radial ports for different purposes e.g., mounting electrodes, feeding argon gas into the chamber, pumping etc..  Two stainless steel circular discs of 100 mm diameter are used as live cathodes, which are kept 30~mm apart from each other. To avoid unnecessary DC glow, the cathodes are covered by teflon caps from the back. A ring shaped grounded anode (100 mm outer diameter, 95 mm inner diameter and 5 mm thickness) is placed symmetrically in between the parallel plate cathodes. Magnetic field is produced by a pair of Helmholtz coils placed symmetrically relative to the centre of the vacuum chamber. The average distance between the two coils is kept 100 mm, which is equal to the radius of the coil. According to the experimental requirements, the magnetic field is varied from 0 to 150 Gauss by increasing the current thorough the coil. The magnetic coils are placed in such a way that the the vacuum electric and magnetic fields become parallel to each other near the cathode as shown in Fig.~\ref{figure1}(b). \par
The chamber is evacuated by using a rotary pump at the base pressure of 10$^{-3}$ torr. The working pressure is set in the range of 0.076 to 0.11 torr by feeding argon gas into the chamber with the help of a gas-dosing valve. A DC discharge is produced between the live cathodes and the grounded anode using a power supply (rating of 600 V and 1.5 A), which is operated in constant voltage mode. In the experiments, the discharge voltage is varied over the range of 250-400V, whereas the discharge current changes accordingly. The discharge voltage is measured by using a digital voltmeter whereas the current is measured precisely by connecting a sensing resistor in the circuit. A tungsten cylindrical Langmuir probe (tip length of 4 mm and diameter of 250 micron) is used to measure the plasma potential and floating potential oscillations near to the anode ring. The probe support is made up of ceramic tube, which is fitted in a stainless steel rod. The grounded anode is used as a reference electrode for Langmuir probe measurements. 
\section{Experimental results and discussion}
\label{sec:results}
\subsection{Magnetic field effect on breakdown voltage}
\begin{figure*}[!]
\includegraphics[width=0.9\textwidth]{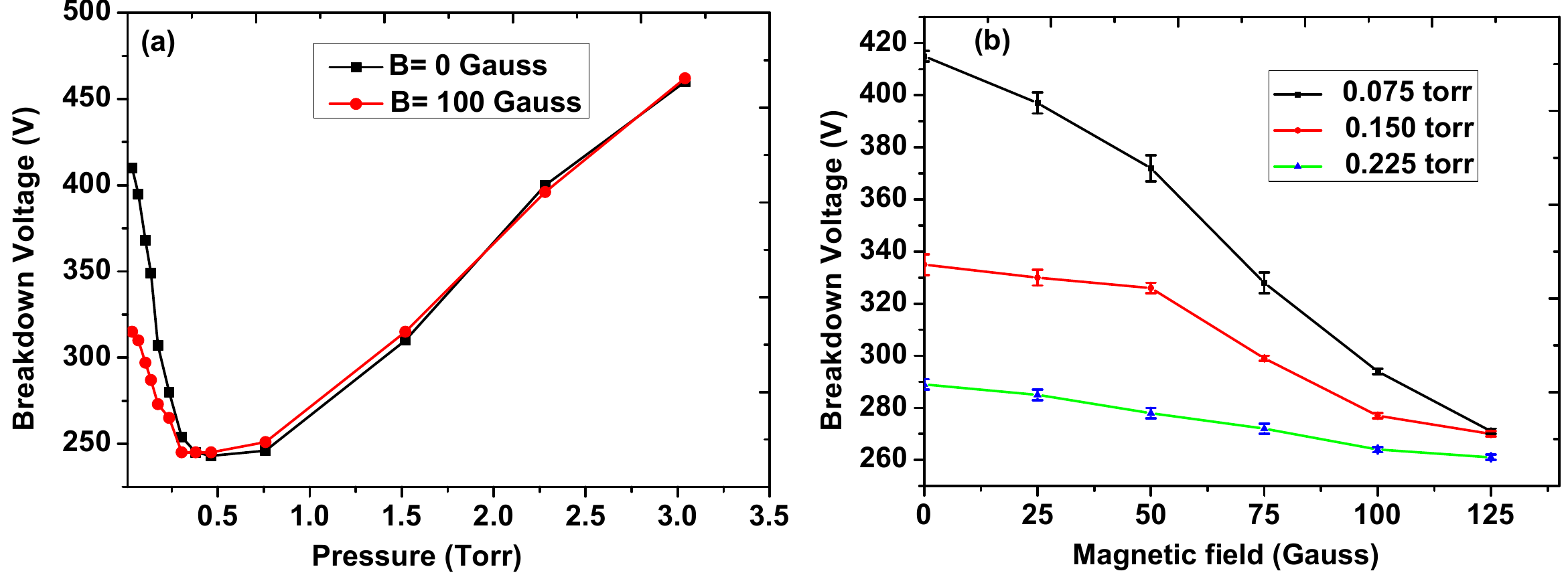}
\caption{Variation of breakdown voltages (a) with background pressure in absence and presence of magnetic field and (b) with magnetic field for different values pressure. The measurement errors are within $\pm5\%$.}
\label{figure2}
\end{figure*}

Initially the effect of magnetic field on breakdown voltage is studied in order to ascertain the pressure range in which the magnetic field influences the discharge properties of the DC glow discharge plasma. For this present set of experiments, the inter-cathode distance is kept around 30 mm and the anode is placed at an equal distance in between two cathodes. The breakdown voltage is measured over a wide range of argon gas pressures (from 0.035~torr to 3.5~torr) in absence and presence of external magnetic field of 100 Gauss as shown in Fig.~\ref{figure2}(a). It is clear from the figure that at a lower pressure (less than 0.2~torr), the magnetic field helps the gas to breakdown at lower voltages. On the other hand, magnetic field has no role on the breakdown voltage if the gas pressure becomes higher than 0.4~torr. \par
%#######################################################################
%########################################################################
In presence of magnetic field, the residence time of the electrons increases, as a result the lateral diffusion of the electrons reduces and leads to a lower breakdown voltage. The electron-neutral collision frequency at higher pressure regime (~ 0.5~GHz at 0.4~torr) exceeds the electron gyration frequency (i.e. 280 MHz at 100 Gauss) and hence the electrons remain un-magnetised. It manifests to an insignificant difference in breakdown voltage curves in presence and absence of magnetic field as shown in fig.~\ref{figure2}(a) However, below 0.4~torr pressure, electron gyration frequency becomes comparable or more than the electron-neutral collision frequency and the effect of magnetic field become significant. Based on this observation, the effect of magnetic field on the breakdown voltage at three different pressure values (0.075, 0.15 and 0.225~torr) is studied and the results are shown in fig. ~\ref{figure2}(b). It is observed that for all these pressure values, the breakdown voltage decreases with the increase of applied axial magnetic field in the range of 25-125~Gauss. This effect turns more pronounced at lower pressure (below 0.225~torr). Breakdown phenomenon of such type of glow discharges is also investigated by Toader et al.\cite{toader2004}, which is in good agreement with our observations.  Therefore, the range of operating pressure from 0.075~torr to 0.11~torr is chosen in our experiments to study the effect of magnetic field on the hysteresis and associated self-excited oscillations.
%%%%%%%%%%%%%%%%%%%%%%%%%%%%%%%%%%%%%%%%%%%%
\subsection{Hysteresis in current--voltage characteristic}
\begin{figure*}[!]
\includegraphics[width=0.9\textwidth]{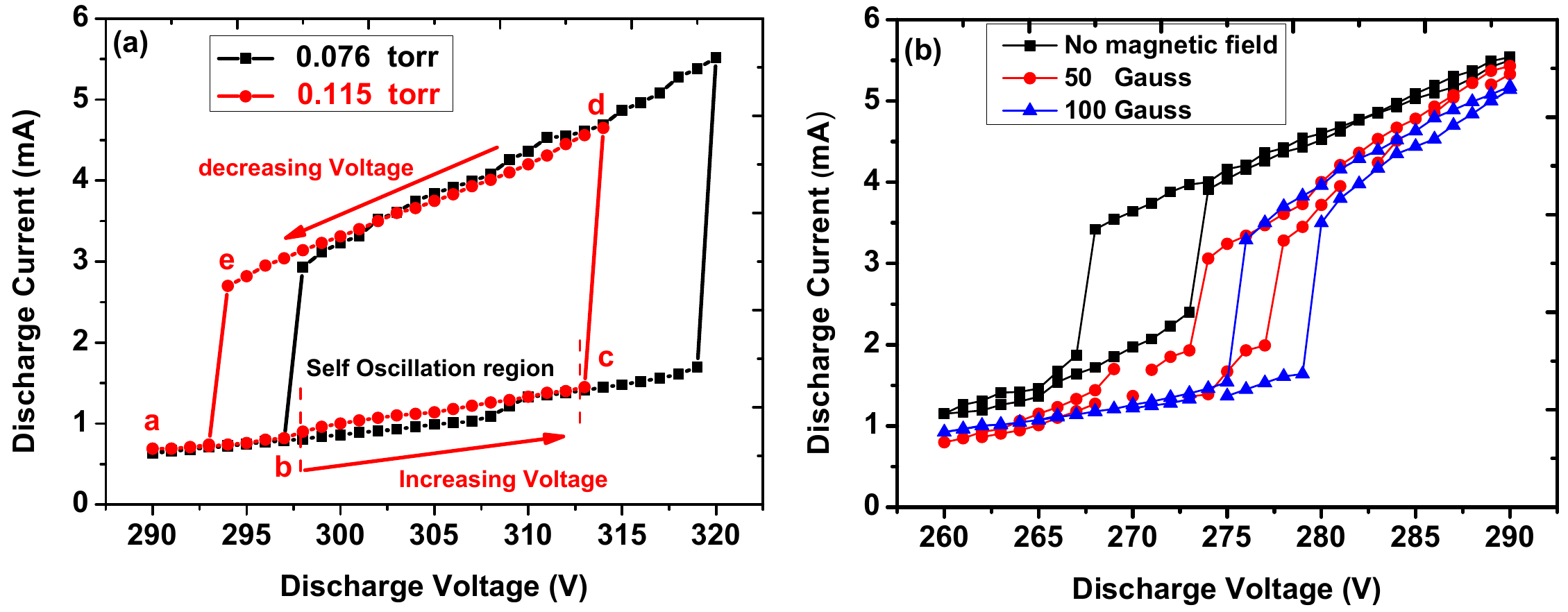}
\caption{Discharge current ($I_d$) voltage ($V_d$) characteristics for different (a) pressures and (b) magnetic fields.The measurement errors are within $\pm5\%$.}
\label{figure3}
\end{figure*}

We then investigate the current-voltage ($I_d-V_d$) characteristics near the breakdown voltage where the discharge current is very low ($\sim$few~mA). Fig.~\ref{figure3}(a) shows $I_d-V_d$ characteristics at two operating pressures (0.075 and 0.11~torr) in absence of external magnetic field. With the increase of discharge voltage, sudden transition to high current mode (HCM) is observed (see point \lq c' in Fig.~\ref{figure3}(a)). However, while decreasing the discharge voltage, the sudden transition back to low current mode (LCM) happens at comparatively lower discharge voltage (see point \lq e' in Fig.~\ref{figure3}(a)), which looks like a hysteresis loop seen in other phenomena. In addition, floating potential is seen to fluctuates at LCM regime of the hysteresis loop (region between \lq b' and \lq c' of Fig.~\ref{figure3}(a)). It is also observed that at low pressure, the voltage required for this transition in discharge current is more. This transition and corresponding hysteresis in the discharge current can be explained as follows:\par
%#######################################################################
%########################################################################
The LCM is generally seen in anode glow of a DC glow discharge plasma. The anode glow/spot occurs when the current to the anode becomes too small to maintain the discharge. This small amount of current can be attributed due to the smaller size of anode or/and due to the restriction of electrons reaching to anode. In order to maintain flux balance, an additional ionisation process begins near the anode by frequent collisions of electrons with neutrals\cite{song1991}. In LCM, ionization takes place mainly near the anode where the energy of the accelerated electrons is higher than the ionization energy of the argon gas atoms. These newly produced ions generated in this process rush towards the cathode. If number of ions created by ionization in anode glow region are smaller than that of ions flowing towards the cathode, the LCM state sustains (see a--c in Fig.~\ref{figure3}(a)). If ion production exceeds, positive charge density in the anode glow region increases. Due to lower mobility of ions, positive potential also increases near the anode, which attracts more electrons from the cathode and leads to more ionization. As a result, anode glow region grows and expands towards cathode. This cascade effect brings a sudden transition (c--d transition) to HCM from LCM regime. The reverse transition from HCM to LCM (point \lq e' in fig.~\ref{figure3}(a)) normally takes place in much lower discharge voltage. This leads to a hysteresis loop as shown in Fig.~\ref{figure3}(a).  The figure essentially signifies that the ionization volume during HCM is much larger than that of LCM. Transition from HCM to LCM occurs only if the ion loss rate overcomes the ion production rate. \par
 In addition to the above investigation, we have studied the hysteresis characteristics for different magnetic fields, which is depicted in Fig.~\ref{figure3}(b). The figure clearly indicates that the transition voltage increases with the increase in magnetic field whereas the total width of the hysteresis loop decreases. It is to be noted that in presence of magnetic field, due to restricted movement of the electrons and ions, the ionization volume in HCM depends on the magnetic field. Higher the magnetic field ionization volume becomes lesser, which leads to lower discharge current in HCM regime and hysteresis area. In Fig.~\ref{figure3}(a) where magnetic field is absent such phenomena is not observed even for different pressure. \par
Since $I_d-V_d$ hysteresis is affected by axial magnetic field, it is necessary to study the effect of magnetic field in the form of $I_d$--B curve, for particular discharge voltage. In the following sections, we discuss the $I_d$ -- B hysteresis study along with the self Ð excited oscillations observed in the LCM regime.
%#######################################################################
\begin{figure*}[!]
\includegraphics[width=0.9\textwidth]{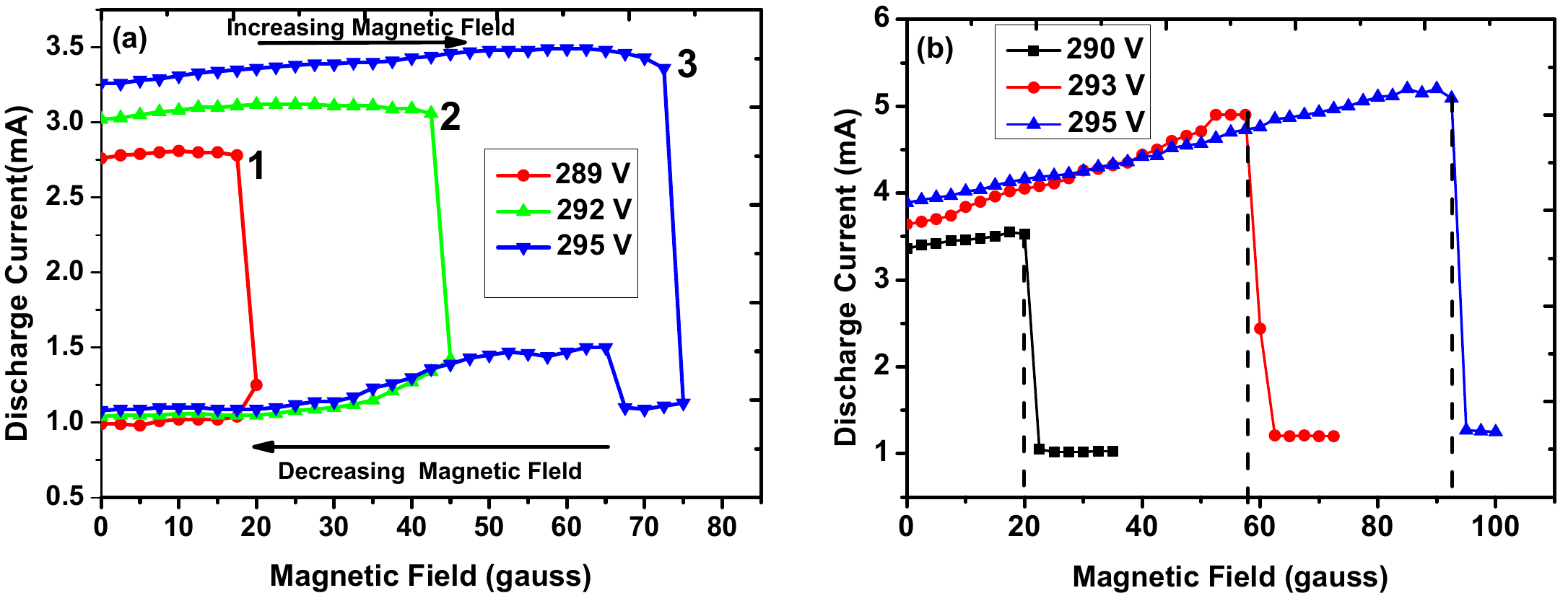}
\caption{Variation of discharge current ($I_d$) (a)  with increased and decreased magnetic field and (b)  with magnetic field after the transition for different discharge voltages. The measurement errors are within $\pm5\%$.}
\label{figure4}
\end{figure*}
%########################################################################
\subsection{Effect of magnetic field on discharge current at HCM of the hysteresis}
\begin{figure*}[!]
\includegraphics[width=0.7\textwidth]{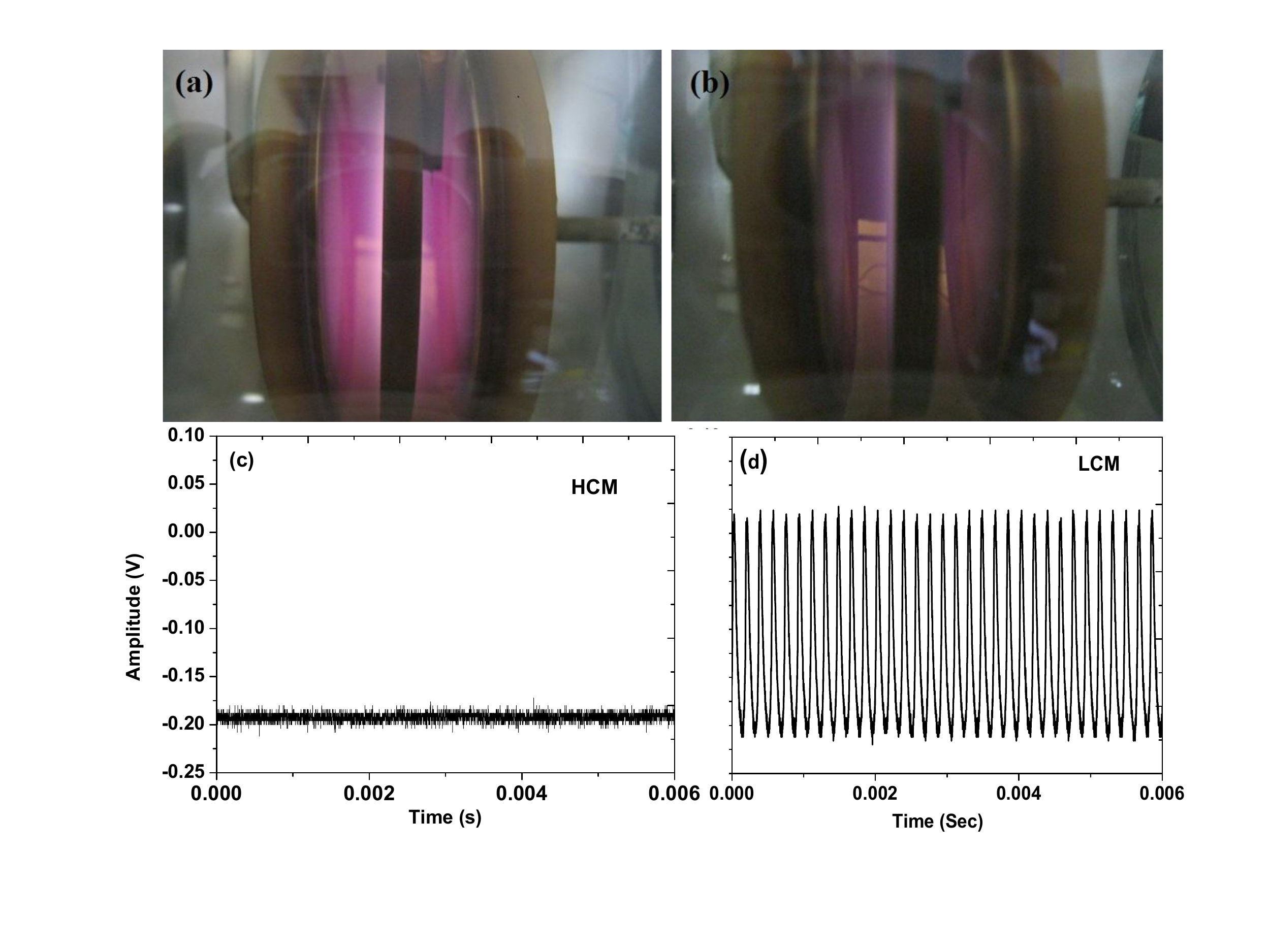}
\caption{Typical glow and floating potential signals (a) $\&$ (c)  in HCM and (b) $\&$ (d)  in LCM regimes, respectively. }
\label{figure5}
\end{figure*}

To examine the effect of magnetic field on the hysteresis characteristics, the discharge current is kept in HCM of hysteresis loop by adjusting the discharge voltage for a given operating pressure of 0.11~torr. Then for a fixed discharge voltage, axial magnetic field is increased in step of 2.5 Gauss by varying the coil current in step of 0.1~A. The change in discharge current is measured and similar exercise is carried out for three different discharge voltages as shown in Fig. 4(a). It is observed that, the critical magnetic field for a particular discharge voltage triggers the transition from HCM to LCM regime and oscillations in the floating potential are observed in LCM regime. The required magnetic field for this transition depends on the initial discharge current. It is clear from Fig.~\ref{figure4}(a) that for 2.75 mA discharge current, magnetic field of ~18 Gauss is sufficient for the transition whereas for $\sim$3.25 mA, 75 Gauss magnetic field is required for the transition. Since the transition from HCM to LCM is linked with electron confinement, hence the discharge voltage and corresponding transition magnetic field follows a linear relationship. In the Fig.~\ref{figure4}(a) all the transition points (1-2-3) are falling on a straight line. The LCM as well as corresponding oscillations in the floating potential continues even after the magnetic field is reduced to the zero. Interestingly the transition from LCM to HCM could not be achieved with further increase in magnetic field rather it always stays in LCM as shown in Fig.~\ref{figure4}(b). It is to be noted that this transition is observed for the discharge voltages within the HCM region of $I_d-V_d$ hysteresis. For higher values of discharge voltage i.e. outside the hysteresis loop, the discharge current increases linearly. The discharge is sustained in high current mode due to the continuous ion bombardment on the cathode surface and subsequently secondary electrons emission from the cathode.  The electrons get accelerated in the cathode sheath and takes part into the ionization process. However in LCM mode, the secondary electron emission reduces from the cathode surface due to low ionization rate. In the case of HCM, the glow covers both the electrodes whereas for LCM the glow confines near the anode ring with low intensity. Fig.~\ref{figure5}(a) shows the typical glow for HCM while Fig.~\ref{figure5}(b) shows the same for LCM. Similarly, Fig~\ref{figure5}(c) shows the floating potential in HCM and Fig~\ref{figure5}(d) shows the floating potential oscillation in LCM.\par
%#######################################################################
%########################################################################
During high discharge current mode, balance between ion production rate and ion loss rate is maintained and higher ionisation volume leads to more discharge current.  However, with increasing magnetic field, the electron gyro-radius reduces which leads to decrease in electron loss rate to the anode. Electron current to the anode is minimised while ion remains un-magnetized for the applied axial magnetic field. The applied axial magnetic field holds the electrons within the anode ring region while ion diffusion is more towards the cathode. Now the anode sheath becomes much dominated by electrons due to confinement by magnetic field. Hence, the discharge is dominated by negative space charge. The simulation study for this type of electrode geometry shows electron space charge in the anode ring\cite{prakash2012}. This negative space charge acts as a potential barrier to low energy electrons while free fall for ions. Higher energy electrons can penetrate this negative barrier. The electrons accelerate up to the energies for which atomic ionisation is more and develops additional ionization near the anode. If the neutral pressure is low (~10$^{-3}$ torr) this low current mode is referred as anode glow mode while if the neutral pressure is high (~10$^{-1}$ torr) this is called as double layer mode\cite{lee1998}. Hence, the observed oscillations are anode glow oscillations in the high pressure discharge. In order to explain the magnetic field effect near anode region, the plasma potential measurement in the anode ring is described in the following section.
%#######################################################################
\begin{figure}[!]
\includegraphics[width=0.45\textwidth]{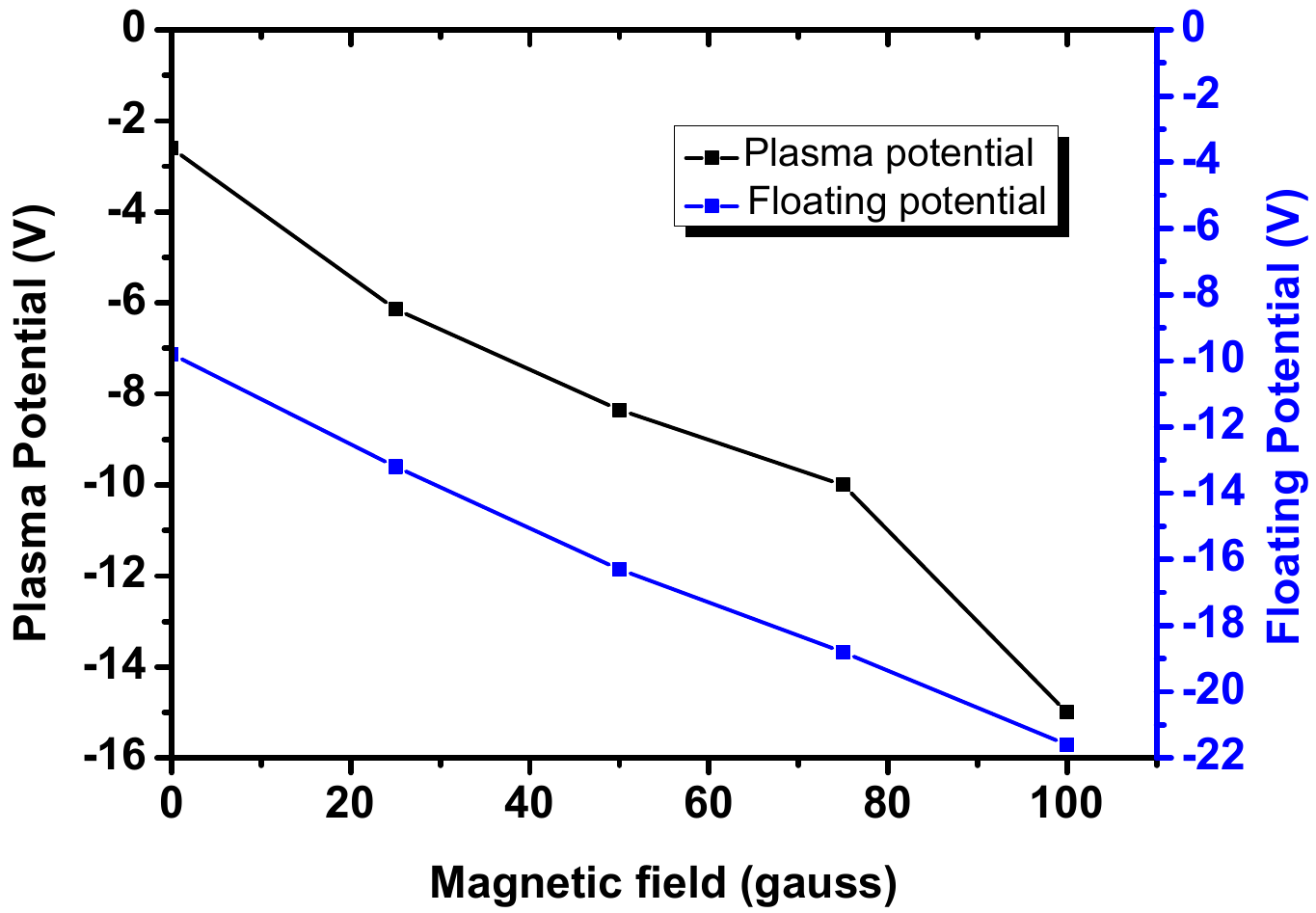}
\caption{Variation of plasma and floating potentials at different magnetic fields. The measurement errors are within $\pm5\%$.}
\label{figure6}
\end{figure}
%########################################################################
\subsection{Effect of magnetic field on plasma and floating potentials in the anode ring}
The Langmuir probe is placed inside the anode ring at 5~mm distance from the top of the anode surface to measure the plasma potential and floating potential. The probe bias voltage with respect to the anode at which the second derivative of probe current becomes zero is considered as plasma potential whereas, the floating potential is a potential at which the net flux of electrons and ions to the probe becomes equal. The plasma and floating potentials are measured for different magnetic field at constant discharge voltage and plotted in Fig.~\ref{figure6}. It is observed that with the increase in magnetic field, the plasma potential becomes more and more negative with reference to the anode. The similar trend is observed for the floating potential measurements. The application of magnetic field suppresses the diffusion transport of the electrons across the magnetic field. At very low discharge voltages in HCM, the cathode fall thickness becomes of the order of inter electrode distance ($\sim$15~mm for 300~V and 3~mA current). In such situation, the available area for electrons to contribute to the discharge current is the inner surface of the anode ring. Because of crossed electric and magnetic fields inside the anode ring, the diffusion of electrons to the anode decreases with the magnetic filed, which results in more negative plasma potential. With the increase in magnetic field, the effective anode area decreases sufficiently as compared to the cathode and hence the anode fall becomes more and more positive and exceeds the ionisation potential. In order to maintain the equality between the ion current to the cathode and electron current to the anode the extra ionisation near the anode takes place and the discharge enters from HCM into LCM. This ionisation phenomenon inside the anode sheath changes the local electric field configuration. These newly generated electrons are then lost easily to the anode while massive ions create a positive space charge region. This generation and collapse of space charge leads to the oscillating behaviour of plasma parameters. The oscillations are also observed in the ion saturation region of the Langmuir probe characteristics when the transition to LCM takes place. 
%%%%%%%%%%%%%%%%%%%%%%%%%%%%%%%%%%%%%%%%%%%%%
\begin{figure*}[!]
\includegraphics[width=.90\textwidth]{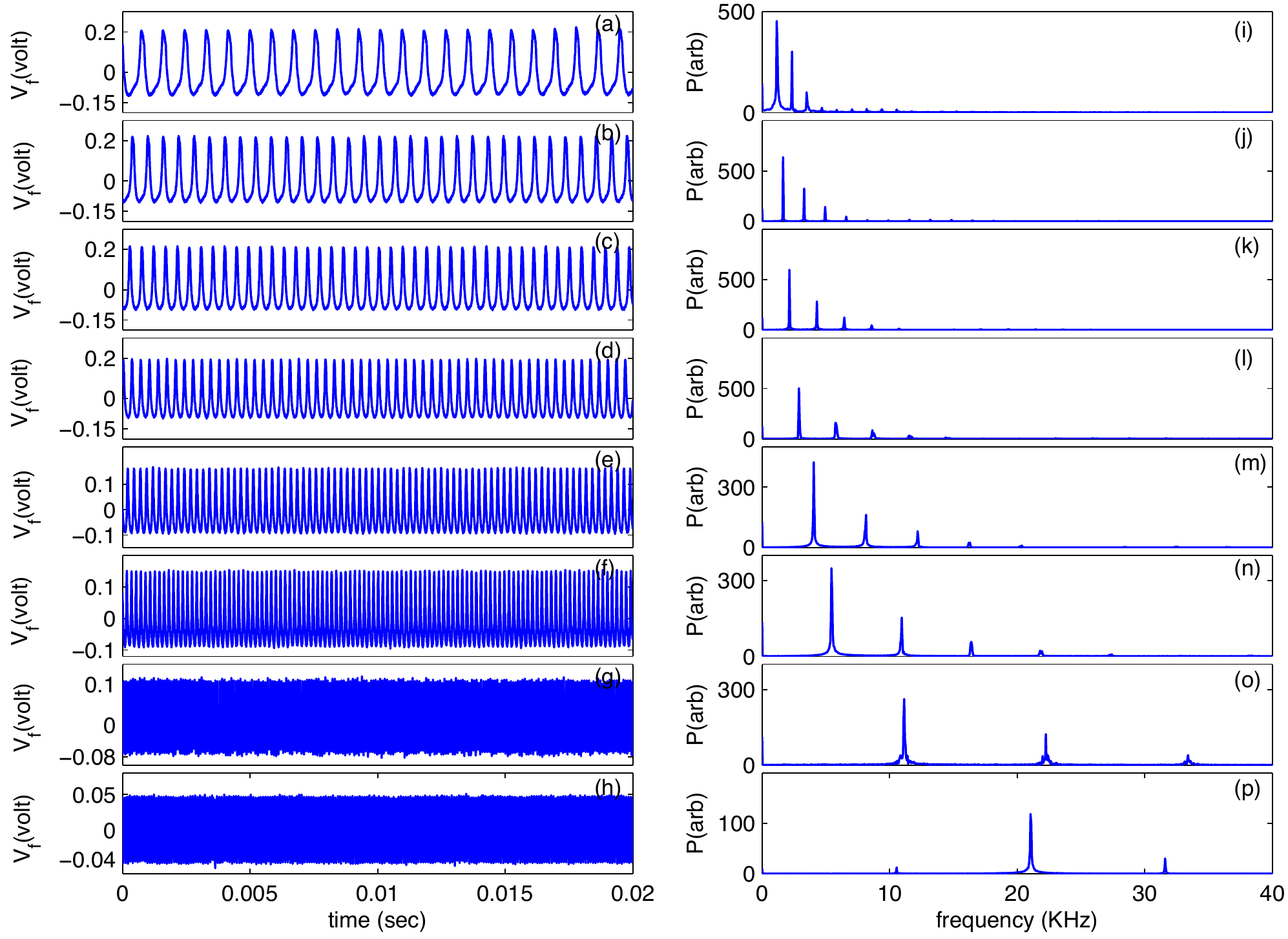}
\caption{Floating potential oscillations (a--h) and their FFTs (i--p) for different magnetic fields: (a) $\&$ (i) 10~Gauss, (b) $\&$ (j) 20~Gauss, (c) $\&$ (k) 30~Gauss, (d) $\&$ (l) 40~Gauss, (e) $\&$ (m) 50~Gauss, (f) $\&$ (n) 60~Gauss, (g) $\&$ (o) 87.5~Gauss and (h) $\&$ (p) 97.5~Gauss. }
\label{figure8}
\end{figure*}
%########################################################################
\subsection{Effect of magnetic field on the floating potential oscillations in LCM}
\begin{figure*}[!]
\includegraphics[width=0.90\textwidth]{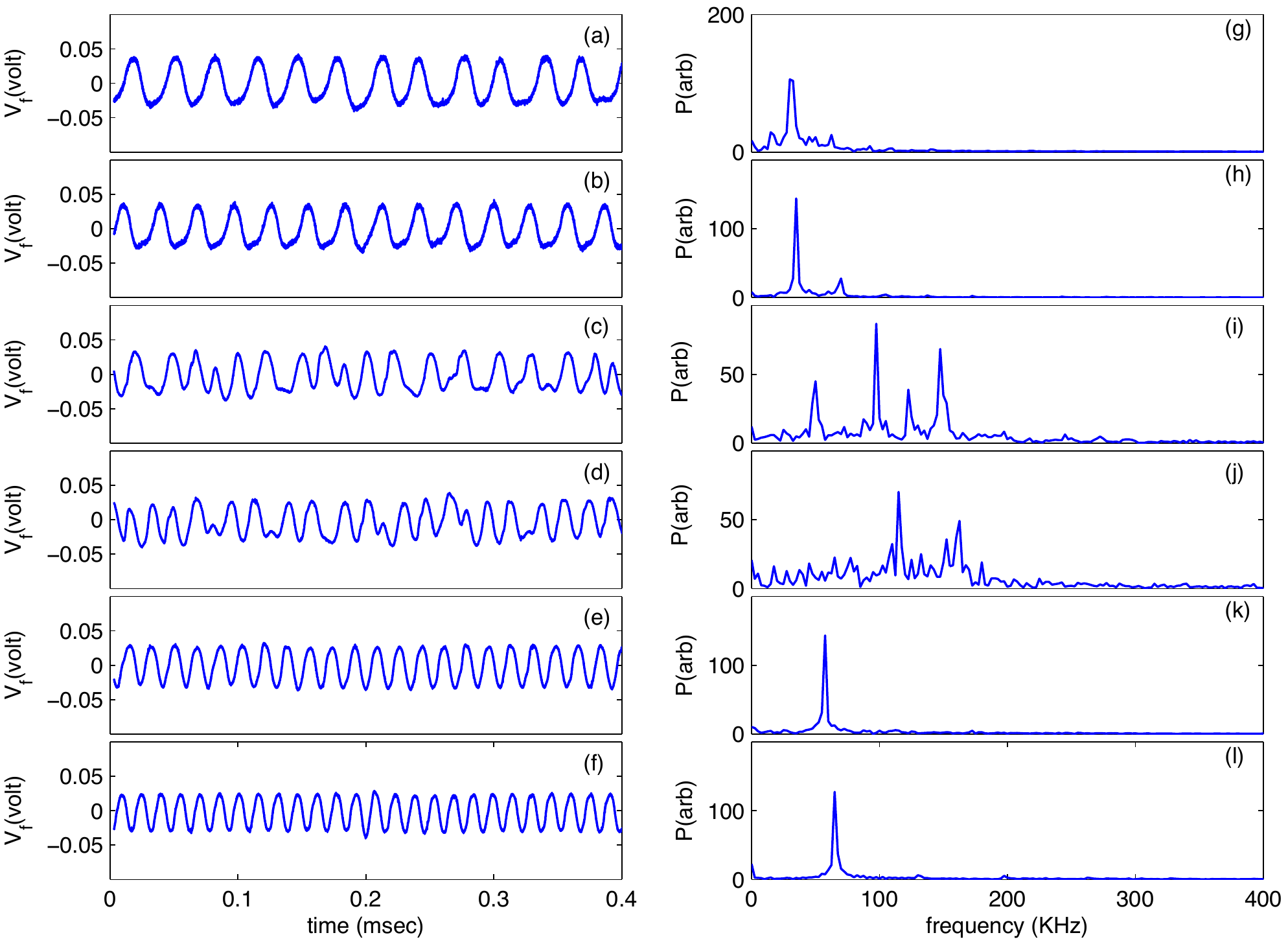}
\caption{Floating potential oscillations (a--f) and their FFTs (g--l) for different magnetic fields: (a) $\&$ (g) 100~Gauss, (b) $\&$ (h) 107.5~Gauss, (c) $\&$ (i) 112.5~Gauss, (d) $\&$ (j) 120~Gauss, (e) $\&$ (k) 125~Gauss and (f) $\&$ (l) 137.5~Gauss }
\label{figure9}
\end{figure*}

As discussed in previous section, LCM is associated with oscillations in discharge parameters e.g. in floating potential and discharge current. The effect of magnetic field on these oscillations is studied over a wide range of magnetic fields. For these studies, the discharge current (0.95~mA) is kept in LCM by adjusting the discharge voltage at 275~V and background pressure at 0.110~torr. The axial magnetic field is then varied from 0 to 137.5~Gauss in step of 2.5~Gauss. The time series of floating potential oscillations (FPO) collected by Langmuir probe at different magnetic field are stored in the oscilloscope and further analysis is conducted with the help of MATLAB software. The floating potential oscillations observed in our experiments are the AC components ($\sim$few mV) that is superimposed on DC level ($\sim$few volts).  Fig.~\ref{figure8}(a)--(h) depicts FPOs, whereas Fig.~\ref{figure8}(i)--(p) represents the corresponding FFT plots, when the magnetic field was increased from 10 to 97.5~Gauss. Similarly, Fig.~\ref{figure9}(a)--(f) shows FPOs, and corresponding FFT plots are given in Fig.~\ref{figure9}(g)--(l) for the increased values of magnetic field from 100 to 137.5~Gauss. At 10~Gauss of magnetic field, the floating potential oscillates at very low frequency $\sim$1.0~KHz.  With increasing magnetic field (see Fig.~\ref{figure8}(i)--(p)) up to 97.5~Gauss, the frequency of oscillation increases. Further increasing of magnetic field from 100 to 107.5~Gauss (see Fig.~\ref{figure9}(g)--(h)), the frequency of oscillations suddenly increases. At 112.5--122.5~Gauss, oscillations become abruptly chaotic as shown in Figs.~\ref{figure9}(c)--(d). It is also observed in FFT analysis (see Fig.~\ref{figure9}(i)--(j) that the dominant peaks are accompanied by smaller peaks, which indicates that the oscillations are indeed chaotic.  With further increasing  magnetic field from 125.5 to 137.5~Gauss, ((see Fig.~\ref{figure9}(e)--(f) the chaotic nature of oscillations resets its periodicity. Therefore, it can be mentioned that with increasing the magnetic field, time series data gradually changes from periodic--chaotic--periodic oscillation. Abrupt changes in oscillations as shown in Fig.~\ref{figure9}(a)--(d) are considered as the transition region. \par
\begin{figure*}[!]
\includegraphics[width=0.80\textwidth]{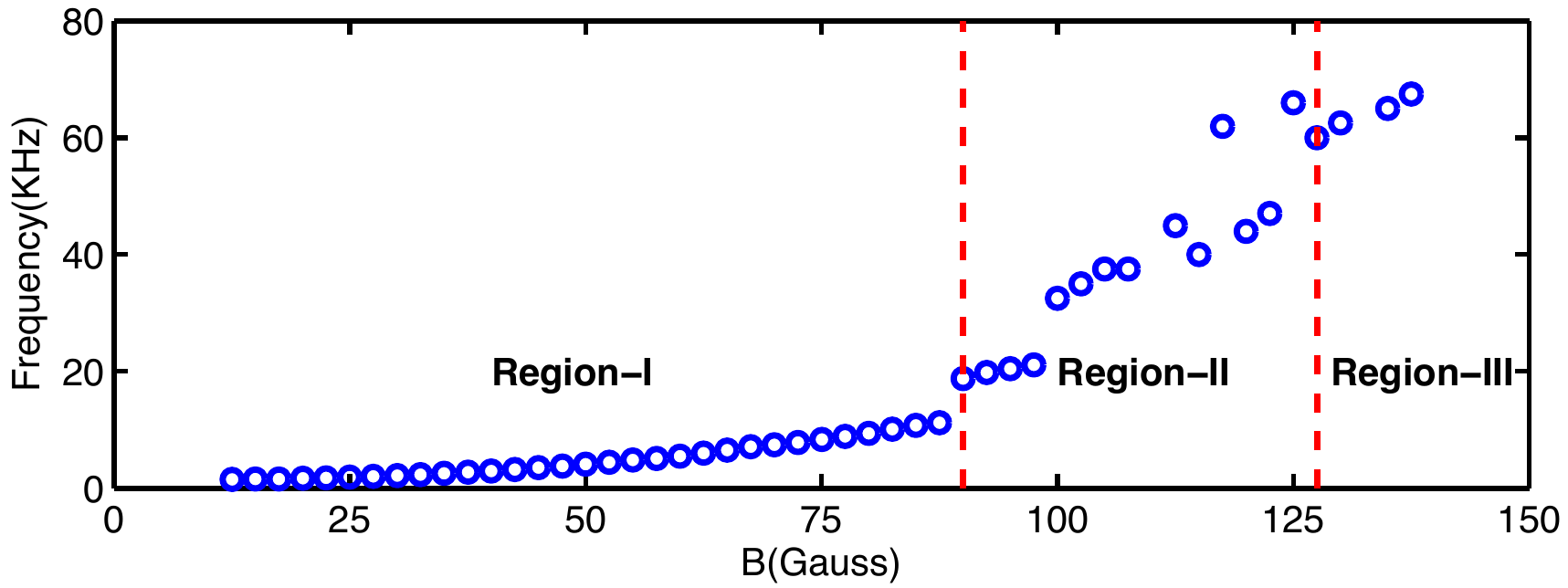}
\caption{Dominated frequencies obtained from FFT plots for different magnetic fields. The measurement errors are within $\pm5\%$.}
\label{figure10}
\end{figure*}

 
Fig.~\ref{figure10} shows the fundamental frequencies obtained from the power spectrum as a function of magnetic field. For a magnetic field of 12.5~Gauss, the frequency of oscillations is observed to be $f\sim1.50$~KHz, with further increasing in magnetic field up to $\sim$90~Gauss, frequency of the oscillation is found to increase linearly (see Region-I). As continuing increment of the magnetic field of 90-127.5~Gauss (chaotic region, Region-II), oscillations loose its periodic increment of the frequency. For further increase in magnetic field from 127.5 to 137.5~Gauss, oscillations regain its periodicity as shown in Region-III.  \par
It is worth mentioning that the range of observed frequency (1--15~KHz) of periodic oscillations always lies below the ion plasma frequency ($\sim$500~KHz) for the our discharge conditions. It has already been reported that the dominant frequency increases with the applied voltages in case of anode glow mode\cite{lee1998}. Interestingly in our experiments, we have observed the similar trend of frequency for increasing magnetic field. In the low discharge current mode, anode glow/spot is observed in the anode ring, where the radial electric field is always perpendicular to axial magnetic field. As a result, the radial electron transport across the magnetic field decreases with the increase in magnetic field. The anode region becomes much dominated by the electron population, which helps to increase the anode fall voltage. Hence, the ion production in the anode sheath increases due to the increase in anode fall and therefore the frequency of oscillation increases. The rising part of the wavefront of Fig.~\ref{figure5}(d) corresponds to the enhancement of ion space charge with time and then it decreases due to loss of ions. With further increase in magnetic field, the rates of creation of ion space charge region and ion loss increases, which results in the increase of oscillation frequency as shown in Fig.~\ref{figure10}. When the magnetic field is increased beyond 90~Gauss the nature of oscillations in the floating potential changes drastically and linear oscillation becomes nonlinear in nature. 
%#######################################################################
%########################################################################
%%%%%%%%%%%%%%%%%%%%%%%%%%%%%%%%%% 
%#######################################################################
%########################################################################
%#######################################################################
\begin{figure}[!]
\includegraphics[width=0.450\textwidth]{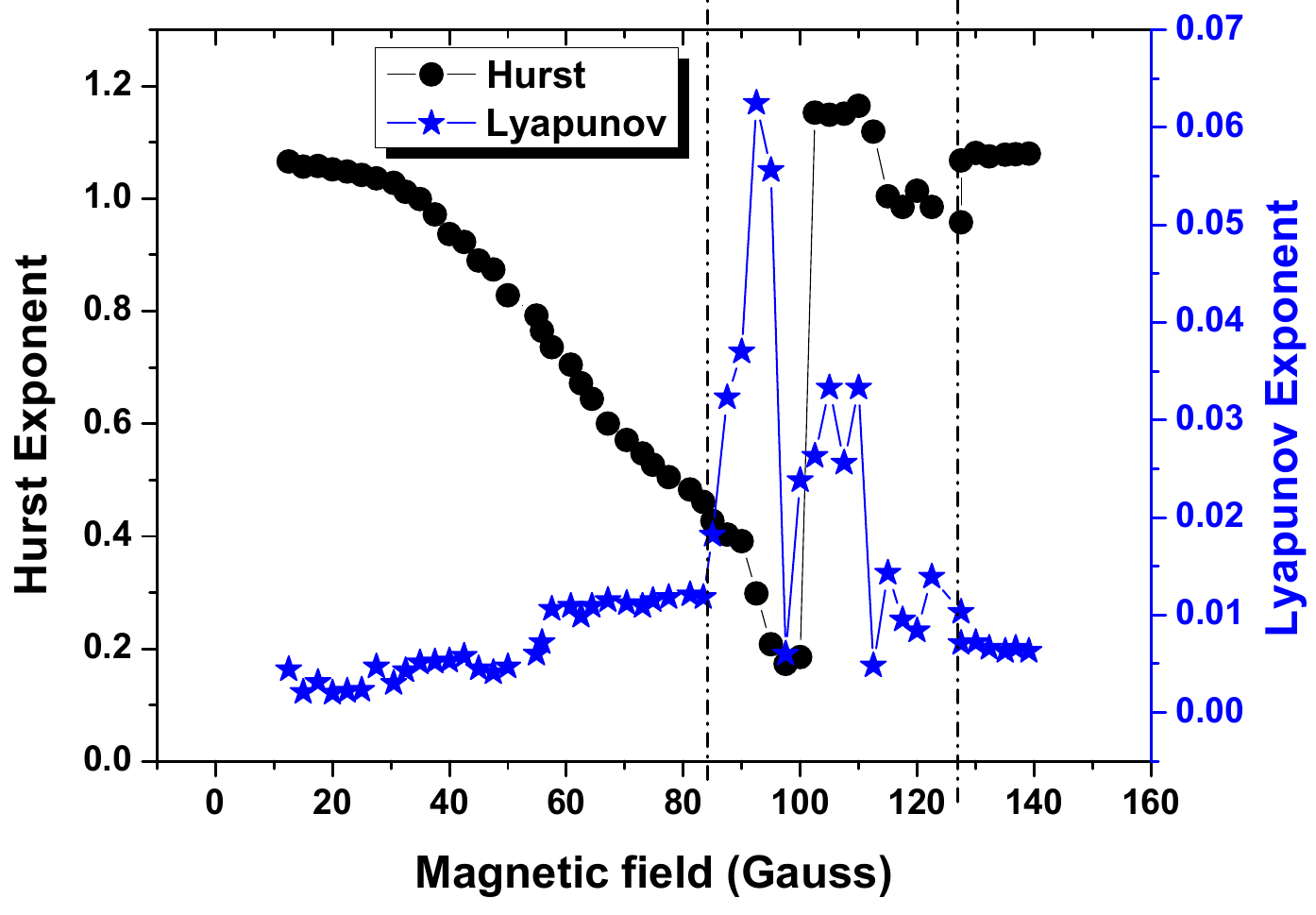}
\caption{Hurst exponent and Lyapunov exponent for different magnetic fields. The measurement errors are within $\pm5\%$}
\label{figure11}
\end{figure}
%########################################################################
\subsection{Nonlinear time series analysis}
It is known from the nonlinear analysis of the oscillation, that the plasma exhibits quasiperiodic--chaotic--quasiperiodic transition, periodic oscillation to chaotic, period doubling, period subtraction etc.\cite{gopi2015,cheung1987,cheung1988,qin1989}. To understand the nature  of oscillations and the corresponding possibility of its predictability, the oscillations of the floating potential are quantified using Hurst Exponent (HE) R/S statistics\cite{mandelbrot1969,adhikary2006} and Largest Lyapunov Exponent (LLE)\cite{kantz1997}. The algorithm of HE and LLE are described elsewhere\cite{nurujiaman2007,nurujiaman2007a}. Chaotic dynamical systems are sensitive to initial conditions and exhibit the exponential divergence in the phase space. The divergence can be calculated by using Lyapunov Exponent (LE), which quantifies the nature of oscillations: positive LE indicates the presence of chaotic nature, whereas  nearly equal to zero and negative LE values indicates the pure periodic. The Hurst exponent is used as a measure of long range correlation in the data. These two methods show the evaluation of transition from regular anticipated plasma oscillation to unpredictable random chaos. \par
If the value of H exponent lies in between 0.5 and 1 then it indicates the persistence, while a value of H exponent equals to 0.5 indicates an uncorrelated process whereas, if its value lies between 0 and 0.5 then it indicates anti-persistence process. Fig.~\ref{figure11} shows the Hurst exponent and largest Lyapunov exponent with increasing magnetic field. It is observed that the Hurst exponent is $\sim$1 at 12.5~Gauss of magnetic field and further increasing the magnetic field up to 100~Gauss Hurst exponent continuously drops from $\sim$1 to $\sim$0.15. This indicates that transition of the oscillation has taken place from persistence to antipersistence.  Further increase in magnetic field, antipersistence is presented in the oscillations and at 127.5 of magnetic field persistence into the oscillations is observed. LLE plot in Fig.~\ref{figure11} shows the exactly opposite trend to the HE and representing that magnetic field from 90 to 127.5 Gauss is transition region. LLE, HE and dominated frequency plots are well supported to each other. In addition to the earlier observations\cite{sharma2013a}, it is seen in our experiments that the nonlinear oscillations in plasma has also a dependency on external magnetic field. 
\section{conclusion}\label{sec:conclusion}
A detailed study on hysteresis phenomenon in discharge current and voltage of a reflex plasma source consisting of modified hollow cathode in presence of an external magnetic field is reported. The discharge current-voltage characteristic shows two stable regimes, a high discharge current regime and a low discharge current regime. The low current regime exhibits the characteristics similar to the anode glow mode, which shows self-oscillations in floating potential at a particular discharge condition.  The effect of external magnetic field on the discharge current is also studied when the discharge is operated within the hysteresis loop. It is observed that the low current mode can be obtained by increasing the magnetic field at low discharge voltages. Applied axial magnetic field alters the discharge current mode from oscillation free to oscillating mode due to restriction of electron transport to the anode. Self-oscillations depends upon discharge voltage, pressure and magnetic field. It is noticed that the frequency of floating potential oscillations increases with the magnetic field. The observed mode transition and oscillations is the effect of anode glow formation near the anode at low discharge current. The nonlinear analysis of these oscillations shows that the periodic to chaotic and again to periodic transition occurs when the magnetic field is increased beyond a threshold value. The knowledge obtained from the nonlinear analysis may have a potential for the application of different control parameters of such experiment through feedback control mechanism.
%+++++++++++++++++++++++++++++++++++++++++
%\section*{References}

\end{document}